\documentclass[journal]{IEEEtran}

\usepackage{url,hyperref,cite,amssymb,amsmath,graphicx,hyperref}
\usepackage{tabularray}
\UseTblrLibrary{booktabs}
\UseTblrLibrary{varwidth}
\usepackage{graphicx}

\begin{document}

\title{Fast FullSubNet: Accelerate Full-band and Sub-band Fusion Model for Single-channel Speech Enhancement}

\author{
Xiang Hao and Xiaofei Li$^{*}$
\thanks{* Corresponding Author}
\thanks{Xiang Hao is now with The Hong Kong Polytechnic University, Hong Kong SAR, China, (e-mail: haoxiangsnr@gmail.com)}
\thanks{Xiaofei Li is with Westlake University and with Westlake Institute for Advanced Study, Hangzhou, China, (e-mail: lixiaofei@westlake.edu.cn)}
}

\markboth{Journal of \LaTeX\ Class Files, Vol. 14, No. 8, August 2015}
{Shell \MakeLowercase{\textit{et al.}}: Bare Demo of IEEEtran.cls for IEEE Journals}
\maketitle

\begin{abstract}
FullSubNet is our recently proposed real-time single-channel speech enhancement network that achieves outstanding performance on the Deep Noise Suppression (DNS) Challenge dataset. A number of variants of FullSubNet have been proposed, but they all focus on the structure design towards better performance and are rarely concerned with computational efficiency. For many speech enhancement applications, a key feature is that system runs on a real-time, latency-sensitive, battery-powered platform, which strictly limits the algorithm latency and computational complexity. In this work, we propose a new architecture named Fast FullSubNet dedicated to accelerating the computation of FullSubNet. 
Specifically, Fast FullSubNet processes sub-band speech spectra in the mel-frequency domain by using cascaded linear-to-mel full-band, sub-band, and mel-to-linear full-band models such that frequencies involved in the sub-band computation are vastly reduced. After that, a down-sampling operation is proposed for the sub-band input sequence to further reduce the computational complexity along the time axis. Experimental results show that, compared to FullSubNet, Fast FullSubNet has only 13\% computational complexity and 16\% processing time, and achieves comparable or even better performance. Code and audio samples are available at \href{https://github.com/Audio-WestlakeU/FullSubNet}{https://github.com/Audio-WestlakeU/FullSubNet}.
\end{abstract}


\IEEEpeerreviewmaketitle

\section{Introduction}

\IEEEPARstart{S}{peech} enhancement aims to improve speech intelligibility and perceptual quality in noisy environments~\cite{loizou_speech_2013}.
Recent Deep Noise Suppression (DNS) Challenge~\cite{reddy_interspeech_2020, reddy_icassp_2021,dubey_icassp_2022} have significantly contributed to advances in the speech enhancement field and fostered many state-of-the-art (SOTA) methods~\cite{hu_dccrn_2020,hao_fullsubnet_2021,li_simultaneous_2021}. Among these methods, FullSubNet~\cite{hao_fullsubnet_2021} attracted broad attention because of its excellent perceptual quality of enhanced speech, especially for the reverberant speech case. Unlike the mainstream methods, which only process the full-band spectra, FullSubNet integrates a full-band model and a sub-band model and performs joint optimization. In FullSubNet, the full-band model extracts global spectral information and long-distance cross-band dependencies. Meanwhile, the sub-band model processes the frequency bands independently and focuses on local spectral patterns and signal stationarity. Experiments show that these two kinds of models are complementary and can be efficiently integrated into one framework.

The fusion scheme proposed by FullSubNet is compatible with other advanced techniques employed in SOTA speech enhancement methods. In the past year, a number of FullSubNet variants~\cite{chen_fullsubnet_2022,dang_dpt-fsnet_2022,chen_lightweight_2022,yuan_dccrn-subnet_2021,xiong_spectro-temporal_2022} have been proposed. DCCRN-SUBNET~\cite{yuan_dccrn-subnet_2021} combines a deep complex convolution recurrent network (DCCRN) and attention gates as an improved full-band model and keeps the sub-band model unchanged. DPT-FSNET~\cite{dang_dpt-fsnet_2022} proposes a dual-path transformer-based full-band and sub-band fusion network. FullSubNet+~\cite{chen_fullsubnet_2022} replaces the LSTM layers in the original full-band model with stacked temporal convolutional network blocks. STSubNet~\cite{xiong_spectro-temporal_2022} uses a novel sub-band network to corporate an efficient spectro-temporal receptive field extractor to achieve simultaneous denoising and dereverberation. Through the sophisticated structure design, these variants achieve remarkable noise suppression performance.
Nevertheless, the computational cost of the full-band and sub-band fusion models remains a blind spot due to the hundreds of times running the sub-band model for processing one signal clip.

This work aims at accelerating the computation of FullSubNet. Instead of straightforwardly seeking to prune or quantize neural networks, we introduce a new architecture named Fast FullSubNet to reduce the complexity caused by the sub-band model while maintaining the speech enhancement performance. Unlike FullSubNet, which works in the linear-frequency domain, this work proposed to work in the mel-frequency domain to largely reduce the number of frequency bands. Mel-frequency presents speech spectra more compactly and meanwhile without losing spectral information in the sense of human auditory perception. Specifically, the linear-frequency spectra are first transformed into the mel-frequency domain, then processed with cascaded full-band and sub-band models following the spirit of FullSubNet. Afterward, an extra mel-to-linear full-band model is added to transform back to the linear-frequency domain, which is similar to the neural vocoders used in recent text-to-speech (TTS) systems~\cite{shen_natural_2018, ping_deep_2022} that  perform a mel-to-linear transformation. Besides, to further reduce the complexity of the sub-band model, we propose to downsample the feature sequence processed by the sub-band model, and then leverage the mel-to-linear full-band model to interpolate the output of the sub-band model. These models and strategies are efficiently integrated together. Experimental results show that Fast FullSubNet reduces the computational complexity and processing time to about 13\% and 16\% of that of FullSubNet. Moreover, Fast FullSubNet achieves comparable or even better speech enhancement performance than FullSubNet due to the use of mel-frequency and post mel-to-linear model. We believe that the design scheme of Fast FullSubNet is also suitable for other full-band and sub-band fusion models \cite{chen_fullsubnet_2022,dang_dpt-fsnet_2022,chen_lightweight_2022,yuan_dccrn-subnet_2021,xiong_spectro-temporal_2022}. 
Code and audio samples are available at \href{https://github.com/haoxiangsnr/FullSubNet}{https://github.com/haoxiangsnr/FullSubNet}.

\section{Method}

\begin{figure}[t]
    \centering
    \centerline{\includegraphics[width=\linewidth]{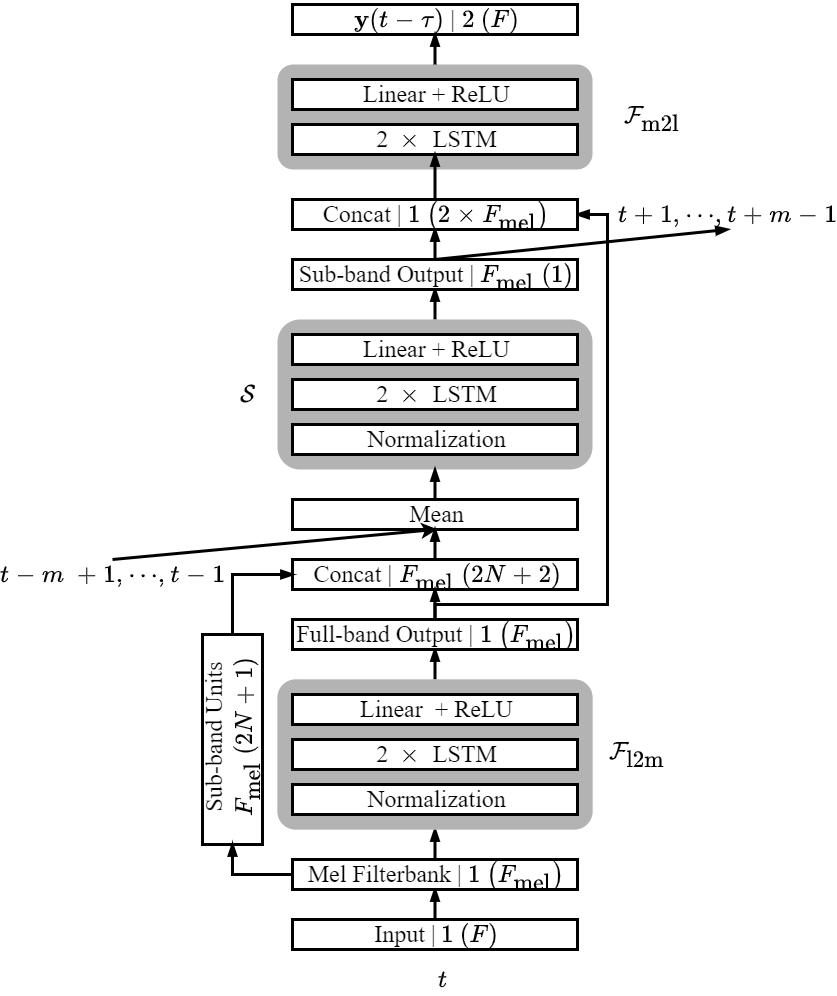}}
    \caption{Diagram of Fast FullSubNet. The right parts of rectangle boxes show feature dimensions, e.g., ``$1~(F)$" represents a $F$-dimension vector. ``$F_\text{mel} (2N + 2)$" denotes $F_\text{mel}$ independent ($2N + 1$)-dimensional vectors.}
    \label{fig:workflow}
\end{figure}

This work processes speech signals in the short-time Fourier transform (STFT) domain. The observed noisy speech signals are given by
\begin{equation}
    x(t,f) = s(t,f) + n(t,f)
\end{equation}
where $x(t,f)$, $s(t,f)$ and $n(t,f)$ represent the complex-valued time-frequency (T-F) bins of noisy speech, noise-free speech (can be the reverberant image signal received at the microphone) and interference noise, respectively, with $t \in [1, \cdots, T]$,  $f \in [0, \cdots, F-1]$, $T$ and $F$ being the time frame, discrete frequency bin,  number of frames and  number of discrete frequencies, respectively.
Note that this work only focuses on the denoising task, which means the purpose of this work is to suppress noise $n(t,f)$ and recover the reverberant speech signal $s(t,f)$.

The proposed Fast FullSubNet reduces the computational complexity of FullSubNet by decreasing the number of frequencies and time frames involved in the sub-band model. 
Figure~\ref{fig:workflow} shows the workflow of Fast FullSubNet, which keeps the motivation and logic of the original FullSubNet unchanged. 
To decrease the number of frequency bands, we first perform speech enhancement in the mel-frequency domain with a linear-to-mel full-band model $\mathcal{F}_{\text{l2m}}$ and a sub-band model $\mathcal{S}$, since the mel-frequency representation of speech is more compact and still informative in terms of human auditory perception. Then, the output of $\mathcal{S}$ is transformed back to the linear-frequency domain with a mel-to-linear full-band model $\mathcal{F}_\text{m2l}$. Each of the three models consists of a two-layer LSTM network.

\subsection{Full-band model $\mathcal{F}_{\text{l2m}}$}


The mel-scale spectral magnitude is first processed by the full-band model $\mathcal{F}_\text{l2m}$ for extracting global spectral information and long-distance cross-band dependencies. 
Formally, the linear frequency signal $x(t,f)$ is transformed to the mel-frequency domain as $x_\text{mel}(t,f), f \in [0, \cdots, F_\text{mel} - 1]$, where $F_\text{mel}$ is the number of mel frequencies. 
The input vector of $\mathcal{F}_\text{l2m}$ at time $t$ is 
\begin{align}
    \mathbf{x}(t) = [x_\text{mel}(t, 0), \cdots, x_\text{mel}(t, F_\text{mel} - 1)]^T  \in \mathbb{R}^{F_\text{mel}}.
\end{align}
The sequence of this feature vector is processed with two layers of LSTM. This full-band model outputs a spectral embedding with the same size as $\mathbf{x}(t)$, namely one hidden unit for each mel frequency. 
This spectral embedding provides complementary information to the following sub-band model.

\subsection{Sub-band model $\mathcal{S}$}
The sub-band model processes the mel frequencies independently, and all frequencies share the same network. The sub-band model predicts clean speech leveraging the signal stationarity and the local spectral pattern of speech. 
Specifically, the input of the sub-band model contains two sources. For one frequency $f$, the first source is the noisy mel-spectra of this frequency and $N$ adjacent frequencies at each frequency side.
The second source is the output of the full-band model at frequency $f$, denoted as $\mathcal{F}_\text{l2m}(\mathbf{x}(t))(f)$. We concatenate these two sources as the input of the sub-band model 
\begin{align}
\label{eq:neighbor}
    \mathbf{x}_\text{sub}(t,f) & =  [ x_\text{mel}(t, f - N), \cdots, x_\text{mel}(t, f),
     \cdots, \notag\\ & x_\text{mel}(t, f + N),
     \mathcal{F}_\text{l2m}(\mathbf{x}(t))(f) ]^T \in \mathbb{R}^{2N + 2}.
\end{align}
The sequence of this feature vector is processed with the same two layers of LSTM for all frequencies.
For each mel frequency, the sub-band model outputs a one-dimensional hidden unit. 

\subsection{Full-band model $\mathcal{F}_\text{m2l}$}

The full-band model $\mathcal{F}_\text{m2l}$ transforms mel-frequency back to linear-frequency. The output of full-band model $\mathcal{F}_\text{l2m}$ and sub-band model $\mathcal{S}$ are concatenated as the input of $\mathcal{F}_\text{m2l}$:
\begin{align}
\label{eq:m2l}
    \mathbf{x}_\text{m2l}(t) = [& \mathcal{F}_\text{l2m}(\mathbf{x}(t))^T, \mathcal{S}(\mathbf{x}_\text{sub}(t,0)), \cdots, \notag \\
    & \mathcal{S}(\mathbf{x}_\text{sub}(t,F_\text{mel} - 1)) ]^T \in \mathbb{R}^{2  F_\text{mel}}.
\end{align}
Two layers of LSTM followed by one linear layer predict the final linear-frequency output. 
The complex-valued Ideal Ratio Mask (cIRM)~\cite{williamson_complex_2016} is taken as the learning target. Denote cIRM as $y(t, f) \in \mathbb{C}$ for one T-F bin. $\mathcal{F}_\text{m2l}$ predicts the real-valued cIRM vector at time $t$ as 
\begin{align}
    \mathbf{y}(t) = [& \text{R}\{y(t, 0)\}, \text{I}\{y(t, 0)\},\cdots, \notag \\
    & \text{R}\{y(t, F-1)\}, \text{I}\{y(t, F-1)\}]^T \in \mathbb{R}^{2 F},
\end{align}
where $\text{R}\{\}$ and $\text{I}\{\}$ denote the real and imaginary parts of complex number, respectively. This mel-to-linear model performs a similar function as the neural vocoders used in TTS~\cite{ping_deep_2022,kaneko_maskcyclegan-vc_2021}, as they both perform mel-frequency to linear-frequency transformation, except that speech enhancement can use the noisy signal phase. 
As shown in Figure~\ref{fig:workflow}, to employ a look-ahead of $\tau$ frames, the target sequence could be set to be delayed $\tau$ frames relative to the input sequence.

\subsection{Sub-band down-sampling}
One key characteristic of speech signals is that the samples are ordered in time, and successive samples are dependent/redundant~\cite{manolakis_statistical_2005}, which means we may process only a part of the samples without performance degradation. To further reduce the computational complexity of the sub-band model, we down-sample the feature sequence of the sub-band model by a factor of $m$. Down-sampling is conducted by non-overlapped averaging the input, i.e., $\mathbf{x}_\text{sub} (t, f)$, for every $m$ frames. For frequency $f$, the down-sampled input is denoted as $\tilde{\mathbf{x}}_\text{sub}(n,f)$, where $n \in [1, \cdots, \lceil \frac{T}{m} \rceil ]$ is the index of down-sampled time frames. When $m=1$, there will be no downsampling.
The down-sampled sequence may lose information employed by the sub-band model, such as the temporal dynamic of local spectral pattern, and thus degrades the quality of the sub-band output.  
 
The output of the sub-band model will be down-sampled accordingly. The down-sampled sub-band output is first copied for $m$ times and then fed to the following full-band model $\mathcal{F}_\text{m2l}$. For this case, $\mathcal{F}_\text{m2l}$ is not only used for mel-to-linear transformation but also for interpolating the sub-band output leveraging the dependence of adjacent frames. Note that, also as the input of $\mathcal{F}_\text{m2l}$, the output of $\mathcal{F}_\text{l2m}$ is not down-sampled, which may alleviate the difficulty of interpolation. 

As shown in Figure~\ref{fig:workflow}, to conduct down-sampling and meanwhile guarantee online processing, at a one-time step, the averaging of input frames only uses the previous $m-1$ time steps, while the output is copied to the future $m-1$ time steps.

\begin{table*}[htp]
\centering
\caption{Performance on DNS Challenge (INTERSPEECH 2020) dataset. For comparison methods, the scores are directly quoted from their original papers, and the missing scores in the original papers are shown as blanks.}  
\resizebox{\textwidth}{!}{%
\begin{tblr}{
    cells={c},
    cell{1}{1} = {r=2, c=1}{c},
    cell{1}{2} = {r=2, c=1}{c},
    cell{1}{3} = {r=1, c=4}{c},
    cell{1}{7} = {r=1, c=4}{c},
    cell{1}{11} = {r=2, c=1}{c},
    cell{1}{12} = {r=2, c=1}{c},
    cell{1}{13} = {r=2, c=1}{c},
    cell{3-9}{1} = {r=1, c=2}{c},
    cell{10}{1} = {r=5, c=1}{c},
    vline{2,3,7,11},
    measure=vbox,
}
\toprule
Method & {Down-sampling \\ factor $m$} & With Reverb & & & & No Reverb & & & & {\# Param \\ (M)} & {MACs \\ (G/s)} & RTF \\ 
\midrule
 &  & WB-PESQ & NB-PESQ & STOI & SI-SDR & WB-PESQ & NB-PESQ & STOI & SI-SDR & & & \\
\midrule
Noisy & &  1.822 & 2.753 & 86.62 & 9.033 & 1.582 & 2.454 & 91.52 & 9.071 & - & - & - \\
\midrule
DTLN~\cite{westhausen_dual-signal_2020} & &  & 2.70 & 84.68 & 10.53 &  & 3.04 & 74.76 & 16.34 & 0.99 & 0.11 & 0.043 \\
DCCRN-E~\cite{hu_dccrn_2020} &  &  & 3.077 & & &  & 3.266 & & & 3.74 & 6.56 & 0.128 \\
Conv-TasNet~\cite{koyama_exploring_2020} &  &  2.75 & & & & 2.73 & & & & 8.68 & 5.97 & 0.659 \\
Sub-band Model~\cite{li_online_2020} & $384 \times 2$ & 2.650 & 3.274 & 90.53 & 14.67 & 2.369 & 3.052  & 94.24 & 16.15  & 1.30 & 21.68 & 0.401 \\
FullSubNet~\cite{hao_fullsubnet_2021} &  & 2.969 & 3.473 & 92.62 & 15.75 & 2.888 & 3.305 & 96.11 & 17.29 & 5.64 & 30.73 & 0.511 \\
\midrule
Full-band Model & & 2.726  & 3.388 &  91.15 & 14.75  & 2.831 & 3.354  & 96.10 & 16.58  &  8.15 & 0.53  & 0.026 \\
\midrule
Fast FullSubNet & 1 & 3.031 & 3.511 & 93.14 & 15.68 & 2.865 & 3.375 & 96.29 & 17.11 & 6.84 & 7.79 & 0.147  \\
                & 2 & 3.016 & 3.497 & 92.96 & 15.85 & 2.808 & 3.353 & 96.11 & 16.98 & 6.84 & 4.12 & 0.082  \\
               & 4 & 2.896 & 3.438 & 92.35 & 15.51 & 2.707 & 3.294 & 95.85 & 16.35 & 6.84 & 2.29 & 0.053 \\
               & 8 & 2.862 & 3.414 & 92.11 & 15.31 & 2.692 & 3.380 & 95.77 & 16.38 & 6.84 & 1.39 & 0.042 \\
            & $+\infty$ & 2.865 & 3.419 & 92.18 & 15.12 & 2.763 & 3.325 & 95.96 & 16.60 & 4.91 & 0.32 & 0.016 \\
\bottomrule
\end{tblr}
}
\label{tab:my_label}
\end{table*}

\section{Experimental Setup}

\subsection{DNS challenge Dataset}

For a fair comparison with FullSubNet, all experiments are conducted on the DNS challenge (INTERSPEECH 2020) dataset. This dataset consists of a clean speech set including about 500-hour clips from 2150 speakers and a noise dataset including over 180-hour clips from 150 classes. The synthesized noisy-clean speech pairs follow the dynamic mixing strategy of FullSubNet.
Before the start of each training epoch, 75\% of the clean speech clips are mixed with randomly selected room impulse responses (RIR) from 
(1) Multichannel Impulse Response Database~\cite{hadad_multichannel_2014} with three reverberation times 0.16 s, 0.36 s, and 0.61 s. (2) Reverb Challenge dataset~\cite{kinoshita_summary_2016} with three reverberation times 0.3 s, 0.6 s and 0.7 s. Then, based on a randomly selected SNR between -5 and 20 dB, the reverberated or no-reverberated speech will be mixed with a randomly selected noise. The total data ``seen" by the model is over 5000 hours after ten epochs of training. We use the test dataset of the DNS Challenge dataset for evaluation, including two categories of synthetic clips, i.e., without and with reverberations. Each category has 150 noisy clips with SNR levels distributed between 0 dB to 20 dB.

\subsection{Implementation Details}

The sampling rate of audio signals is 16000 Hz. STFT uses a 32 ms (512 samples) Hanning window and a 16 ms hop size. The number of mel-frequency bins is set to \textbf{64}. For training, we adopt the Adam optimizer with a learning rate of 0.001. For a fair comparison, same as the parameters mentioned in~\cite{li_online_2020,hao_fullsubnet_2021}, we set output delay $\tau$ to two frames so that the model exploits $16 \times 2 = 32$ ms future information. The sequence length for training is set to the $T=192$ frames (about 3 s). According to preliminary experiments, the number of neighbor frequencies $N$ in Equation~(\ref{eq:neighbor}) is set to 5. 
The three models all consist of two stacked unidirectional LSTM layers and a linear layer. The two full-band models and one sub-band model, have 384/257, 512/512, and 384/384 hidden units for their own two LSTM layers, respectively. This paper uses the Mean Squared Error (MSE) as a loss function.


\section{Experimental Results}

Table~\ref{tab:my_label} shows the experimental results in terms of commonly-used Perceptual Evaluation of Speech Quality (PESQ)~\cite{rix_perceptual_2001}, short-time objective intelligibility (STOI)~\cite{taal_algorithm_2011}), and Scale-Invariant Signal-to-Distortion Ratio (SI-SDR)~\cite{roux_sdr_2019} metrics. Besides, the ``\# Param'' column shows the number of parameters. We also report the Mult-Add calculations (MACs) calculated using the ``torchinfo" tool \footnote{\url{https://github.com/TylerYep/torchinfo}} to show the computational complexity. In addition, real-time factor (RTF), a metric for measuring the inference speed, is also measured on a platform with Intel (R) Core (TM) i7-9700 CPU @ 3.00 GHz and PyTorch 1.12.

\subsubsection{\textbf{Effectiveness of reducing frequencies}}

When the downsampling factor $m$ of Fast FullSubNet is set to one, there is no time downsampling. Compared with the original FullSubNet, decreasing the frequencies for sub-band processing (from 257 to 64) reduces MACs and RTF to about 25\% and 29 \% of that of FullSubNet. Moreover, almost all performance measures are comparable or even better, possibly due to the use of mel-frequency and a post mel-to-linear full-band model. The advantage of using mel-frequency will be more clear later. 

\subsubsection{\textbf{Effectiveness of sub-band downsampling}}
Another set of experiments is conducted with increasing downsampling factors. 
Compared with $m =  1$, $m =2$ achieves a comparable enhancement performance, and MACs and RTF further reduce to about 13\% and 16\% of that of FullSubNet, respectively. This result fits our expectation that the successive time frames are relatively dependent/redundant, and the mel-to-linear full-band model can well interpolate the down-sampled sub-band output.
Decreasing $m$ to 4 and 8, MACs and RTF will be further reduced but at the cost of enhancement performance degradation. Finally, when $m = + \infty$, the sub-band model is totally removed, which achieves similar speech enhancement performance as $m = 8$, which means setting up to $m = 8$, the sub-band model is no longer useful for speech enhancement. 
We would like to note that our preliminary experiments show that the performance degradation caused by sub-band downsampling is related to the number of look-ahead frames, as the look-ahead frames provide more recent information for the interpolation of the output of the sub-band model. This means a smaller/larger number of look-ahead frames will allow using a smaller/larger down-sampling factor, without suffering from performance degradation. 

When the sub-band model is removed, i.e., $m = + \infty$, the network consists of two layers of mel-frequency and two layers of mel-to-linear full-band LSTMs. To testify the use of mel-frequency, we train a Full-band Model (as shown in Table~\ref{tab:my_label}) composed of four layers of 512-dim LSTMs with linear-frequency input and output. It can be seen that the mel-frequency network performs better than the Full-band Model in terms of both speech enhancement performance and computational complexity. The possible reason is that mel-frequency represents speech spectra in a more compact way, which eases the learning of mapping between noisy and clean spectra.

\subsubsection{\textbf{Comparsion with SOTAs}}
We compare Fast FullSubNet with several recent SOTA methods that provide results on the DNS Challenge dataset and open-source their implementation to compare MACs and RTF fairly.
It can be seen that FullSubNet has already outperformed these methods in terms of speech enhancement metrics. 
The proposed Fast FullSubNet with $m=2$ has smaller MACs and RTF than these methods except for DTLN~\cite{westhausen_dual-signal_2020}.
Notably, the flexible strategies for reducing computational complexity provided by Fast FullSubNet are also suitable for other FullSubNet variants.

\section{Conclusion}

This paper proposes a new architecture named Fast FullSubNet to accelerate the computation of FullSubNet by reducing the number of frequencies and time frames involved in the computation of the sub-band model. Experimental results show that compared with the original FullSubNet, Fast FullSubNet achieves comparable or better performance with significantly smaller complexity. Importantly, the flexible strategies for reducing computational complexity provided by Fast FullSubNet are also suitable for other FullSubNet variants.

\bibliographystyle{IEEEbib}
\bibliography{references}

\end{document}